\documentclass[aps,prd,twocolumn,superscriptaddress,preprintnumbers,floatfix]{revtex4-2}
\usepackage{amsmath}
\usepackage{mathrsfs}
\usepackage{slashed}
\usepackage{gensymb}
\usepackage{amsfonts}
\usepackage{mathrsfs}
\usepackage{bbding}
\usepackage{graphicx}
\usepackage{color}
\usepackage{siunitx}
\usepackage{float}
\usepackage[colorlinks=true, linkcolor=blue, citecolor=green]{hyperref}
\usepackage{subfigure}
\usepackage[utf8]{inputenc}
\usepackage{xcolor}

\begin{document}
\title{Angular Resolution of the Search for Anisotropic Stochastic Gravitational-Wave Background with Terrestrial Gravitational-Wave Detectors}
\author{Erik Floden}
%\email[]{flode012@umn.edu}
\affiliation{School of Physics and Astronomy, University of Minnesota, Minneapolis, MN 55455, USA}
\author{Vuk Mandic}
\affiliation{School of Physics and Astronomy, University of Minnesota, Minneapolis, MN 55455, USA}
\author{Andrew Matas}
\affiliation{Max Planck Institute for Gravitational Physics (Albert Einstein Institute), D-14476 Potsdam, Germany}
\author{Leo Tsukada}
%\affiliation{Department of Physics, The Pennsylvania State University, University Park, PA 16802, USA}
\affiliation{Institute for Gravitation and the Cosmos, The Pennsylvania State University, University Park, PA 16802, USA}

%\affiliation{The Pennsylvania State University, University Park, PA 16802, USA}

%\author{Other Author}
%\affiliation{School of Physics and Astronomy, University of Minnesota, Minneapolis, MN 55455, USA}

\date{\today}
\begin{abstract}
We consider an anisotropic search for the stochastic gravitational-wave (GW) background by decomposing the gravitational-wave sky into its spherical harmonics components. Previous analyses have used the diffraction limit to define the highest-order spherical harmonics components used in this search. We investigate whether the angular resolution of this search is indeed diffraction-limited by testing our ability to detect and localize simulated GW signals. We show that while using low-order spherical harmonics modes is optimal for initially detecting GW sources, the detected sources can be better localized with higher-order spherical harmonics than expected based on the diffraction limit argument. Additionally, we discuss how the ability to recover simulated GW sources is affected by the number of detectors in the network, the frequency range over which the search is performed, and the method by which the covariance matrix of the GW skymap is regularized. While we primarily consider point-source signals in this study, we briefly apply our methodology to spatially-extended sources and discuss potential future modifications of our analysis for such signals. 
%Abstract text \cite{LIGOScientific:2019gaw}. \\\\
%\textbf{Keywords}: gravitational wave, stochastic background, spherical harmonics, resolution, other keywords.
\end{abstract}
\maketitle
\section{Introduction}\label{sec:intro}
%\textcolor{blue}{Introductory text...}
Since the first direct detection of gravitational-waves (GW) from the collision of two black holes on September 14, 2015 \cite{gw150914}, the field of gravitational-wave astronomy has been an exciting and relatively new way to probe the far reaches of our universe. The ability to detect gravitational-waves with detectors like the Laser Interferometer Gravitational-wave Observatory (LIGO) and Virgo detectors \cite{2015CQGra..32g4001L,2015CQGra..32b4001A} %\VUK{Add references for LIGO and Virgo detectors, see suggestions by the P\&P} 
has opened up a wide range of questions to investigate. Recent developments include the observation of neutron star-black hole coalescences \cite{Abbott:2021NSBH}, a binary black hole coalescence with a total mass of 150 $M_{\odot}$ \cite{Abbott:GW190521}, and a coalescence between a black hole and a 2.6 $M_{\odot}$ object which is either the lightest black hole or the heaviest neutron star discovered in a compact binary system \cite{Abbott:GW190814}. 
%\VUK{The rest of this paragraph needs to be updated}
As of the end of LIGO-Virgo's third observing run (O3), 90 GW event candidates have been detected \cite{2021gwtc3}, and merger rates have been estimated to be between $17.3$ Gpc$^{-3}$ yr$^{-1}$ and $45$ Gpc$^{-3}$ yr$^{-1}$ for binary black holes, between $13$ Gpc$^{-3}$ yr$^{-1}$ and $1900$ Gpc$^{-3}$ yr$^{-1}$ for binary neutron stars, and between $7.4$ Gpc$^{-3}$ yr$^{-1}$ and $320$ Gpc$^{-3}$ yr$^{-1}$ for neutron star-black hole binaries \cite{2021population}.

%and merger rates have been estimated at $23.9^{+14.3}_{-8.6}$ Gpc$^{-3}$ yr$^{-1}$ for binary black holes and $320^{+490}_{-240}$ Gpc$^{-3}$ yr$^{-1}$ for binary neutron stars \cite{Abbott:Populations}. Two coalescences of neutron star-black hole (NSBH) binaries have been detected, resulting in the calculation an NSBH merger rate of $130^{+112}_{-69}$ Gpc$^{-3}$ yr$^{-1}$ \cite{Abbott:2021NSBH}.

While these detections are of signals from individual GW events, it is also possible for a GW background to form as the superposition of many unresolved GW signals \cite{maggiore, Christensen:2018iqi}. Such a background may have contributions which are astrophysical or cosmological in origin. Examples of astrophysical contributions include binary mergers and supernovae \cite{tania,2009MNRAS.398..293M,isotropic2020,o2iso_arxiv,gw150914_stoch,170817_sgwb_implications,o1iso} %\VUK{Add LVK isotropic SGWB paper references here, they also make predictions of CBC SGWB}
while examples of cosmological contributions include GWs generated during the inflationary epoch and in phase transitions in the early universe \cite{O3_phase_transitions,Geller:2018mwu,vonHarling:2019gme} %\VUK{these refs should also be updated, eg see references in 2110.01478}
. Moreover, the GW background is likely to be anisotropic. The potential sources of anisotropy include primordial density fluctuations (e.g. reflected in the distribution of compact binaries throughout the universe), the local distribution of GW sources (e.g. pulsars in the Milky Way galactic plane \cite{talukder2014measuring,mazumder2014astrophysical}), the velocity of the Solar System, and others \cite{Jenkins:2018lvb,Jenkins:2018uac,Jenkins:2018kxc,Cusin:2017mjm,Cusin:2018rsq,Cusin:2019jpv}.
%\VUK{we should add references to the Cusin et al papers and to Jenkins and Sakellariadou papers about anisotropy}

%we would not necessarily expect this background to be identical in all directions. Anisotropies may exist due to the fact that resolvable GW sources are not expected to be distributed isotropically \cite{gwtc-1}, and different background contributions may be identified by their characteristic anisotropies, such as the distribution of millisecond pulsars in the direction of galactic clusters or neutron stars in our galaxy which trace out the galactic plane . 

Advanced LIGO and Advanced Virgo data have been used to search for the anisotropic stochastic gravitational-wave background (SGWB), producing stringent upper limits on GW energy density across the sky \cite{o3aniso,Abbott:O2_aniso,o1_directional}. Traditionally, diffraction limit arguments were used to assess the angular resolution of these searches. In this paper, we take a closer look at the intrinsic angular resolution of anisotropic SGWB searches, finding that they can surpass then resolution expected based on diffraction limit arguments. In Section II we present the formalism for anisotropic SGWB search. In Section III we present some intuitive arguments on angular resolution limitiations. In Sections IV, V, and VI we present the angular resolution in recovery of simulated individual point sources, simulated multiple point sources, and simulated extended sources, respectively. We offer concluding remarks in Section VII.

\section{Spherical Harmonics Decomposition}\label{sec:shd}
We consider the anisotropic SGWB search in which we decompose the map of the gravitational-wave sky into spherical harmonics components. 
%This search is ideal for identifying spatially extended anisotropies in the SGWB that have smooth spectral behavior. Such a background could result from cosmological sources or astrophysical sources. A cosmological background would result from gravitational-waves left over from the very early universe, analogous to the cosmic microwave background. We expect cosmological backgrounds to have smooth strain power spectra, and in the case of a standard inflationary model, its power spectrum would be constant in frequency. Astrophysical backgrounds, on the other hand, arise from the superposition of many unresolved events such as compact binary mergers, and these backgrounds are expected to have power spectra peaked at particular characteristic frequencies. \cite{sph_methods}
We assume an unpolarized, Gaussian, and stationary SGWB. The most general quadratic expectation value of the GW strain $h_A(f,\Theta)$ of frequency $f$, sky direction $\Theta$, and polarization $A$, that satisfies these assumptions is given by:
\begin{align}\label{eq:strain}
	    \langle h^*_A(f,\Theta)h_{A^{\prime}}(f^{\prime},\Theta^{\prime})\rangle=\frac{1}{4}\mathcal{P}(f,\Theta)\delta_{AA^{\prime}}\delta(f-f^{\prime})\delta(\Theta,\Theta^{\prime})
\end{align}
where ${\mathcal{P}}(f,\Theta)$ gives the spectral and angular distribution of the background \cite{sph_methods}. We assume that ${\mathcal{P}}(f,\Theta)$ can be factored into its separate spectral and angular components
	%\begin{align}
%		\Omega_{\rm{GW}}(f,\Theta)=\frac{2\pi^2}{3H_0^2}f^3 H(f){\mathcal{P}}(\Theta),
%		\label{Eq:omega}
%	\end{align}
	
    \begin{align}
    \label{Eq:p_f_theta}
     \mathcal{P}(f, \Theta) = H(f) \, \mathcal{P}(\hat{\Theta}),
    \end{align}
where $H(f)$ is a dimensionless quantity which we choose to take the form of a power law both for its simplicity and its ability to approximate most interesting SGWB models, 
     \begin{align}
     H(f) =\left(\frac{f}{f_{\rm ref}}\right)^{\alpha -3}.
     \end{align}
Here we use the reference frequency ${f_{\rm ref}} = 25$ Hz. The spectral index $\alpha$ is commonly assumed to take values of 0, 2/3, and 3 corresponding to a cosmological background, CBC background, and a generic flat strain spectrum, respectively. 
We further decompose the angular distribution $\mathcal{P}(\Theta)$ into a basis of spherical harmonic components $Y_{\ell m}$:  
	\begin{align}\label{eq:plm_sum}
		\mathcal{P}(\hat{\Theta})=\sum_{\ell=0}^{\ell_{max}} \sum_{m=-\ell}^{\ell}P_{\ell m}Y_{\ell m}(\hat{\Theta}).
	\end{align}
The goal of the search is therefore to estimate the values of the $P_{\ell m}$ coefficients. To do this, we must first define the cross-correlation spectrum between two detectors $I$ and $J$ at time $t$ and frequency $f$:
	\begin{align}\label{eq:crosscorr}
		C_{IJ}=\frac{2}{\tau}\tilde{s}_I^*(t;f)\tilde{s}_J(t;f).
	\end{align}
Here $\tau$ is the duration of an observation segment, and we use the finite-time Fourier transform of each detector time series data, which contains both the detector noise and the GW signal: $\tilde{s}_I(t;f) = \tilde{n}_I(t;f) + \tilde{h}_I(t;f)$. If we assume that the noise between the detectors is uncorrelated, then the expectation value of the cross-correlation spectrum is:
	\begin{align}\label{eq:crosscorr_quad1}
		\langle C_{IJ}(t;f)\rangle=\frac{2}{\tau}\langle {h}^*_I(t;f){h}_J(t;f)\rangle.
	\end{align}

Equations \ref{eq:strain} and \ref{Eq:p_f_theta} then lead to \cite{sph_methods}:
	\begin{align}\label{eq:crosscorr_integral}
		\langle C_{IJ}(t;f)\rangle=H(f)\int_{S^2} d\hat{\Theta}\gamma(t;\hat{\Theta},f)\mathcal{P}(\hat{\Theta}).
	\end{align}
where $\gamma(t;\hat{\Theta},f)$ is a geometric function that takes into account the response of the detector pair to GW signals given the detectors' relative locations and orientations \cite{sph_methods}:
	\begin{align}\label{eq:overlap}
		\gamma(t;\hat{\Theta},f)=\frac{1}{2}F^A_I(t;\hat{\Theta})F^A_J(t;\hat{\Theta})e^{i2\pi f\hat{\Theta}\cdot(\vec{x}_I(t)-\vec{x}_J(t))/c}%H(f)\int_{S^2} d\hat{\Theta}\gamma(t;\hat{\Theta},f)\mathcal{P}(\hat{\Theta}).
	\end{align}
In the above equation we use the detector response functions $F^A_I(t;\hat{\Theta})$, detector locations $\vec{x}_I(t)$ \cite{sph_methods}, and speed of light $c$. We can substitute Eq. \ref{eq:plm_sum} into Eq. \ref{eq:crosscorr_integral} and integrate over the two-sphere to obtain
	\begin{align}\label{eq:crosscorr_quad2}
		\langle C_{IJ}(t;f)\rangle = H(f) \gamma_{lm}(t;f) \mathcal{P}_{lm},
	\end{align}
where
	\begin{align}\label{eq:gammalm}
		\gamma_{lm}(t;f) = \int_{S^2} d\hat{\Theta} \gamma(t;\hat{\Theta},f) Y_{lm}(\hat{\Theta})
	\end{align}
are purely geometric factors associated with the $IJ$ detector pair and repeated indices are summed over.

We proceed to define a likelihood function for the cross-correlation spectrum given a sky map defined by spherical coefficients $\{\mathcal{P}_{lm}\}$ \cite{o3aniso}:
\begin{align}
    & p(C_{IJ}|\{\mathcal{P}_{lm}\}) \propto  \, {\rm exp} \Big( [C_{IJ}(t;f)- H(f) \, \gamma_{lm} (t;f) \, \mathcal{P}_{lm}]^* \nonumber \Big. \\ 
    & \Big. N^{-1}_{ft,f't'}  [C_{IJ}(t';f')- H(f') \, \gamma_{l'm'}(t';f') \, \mathcal{P}_{l'm'}] \Big)
\end{align}
where $N_{ft,f't'}$ is the covariance matrix of $C_{IJ}(t;f)$ given by \cite{romano2017detection}:
\begin{align}
    N_{ft,f't'}=\delta_{tt'}\delta_{ff'}P_I(t;f)P_J(t;f),
\end{align}
and $P_I(t;f)$ is the one-sided power spectrum of the data from detector $I$. The spherical harmonic coefficients that maximize this likelihood function are given by: 
	\begin{align}\label{eq:CleanMap}
		\hat{P}_{lm}=\sum_{l'm'} \big(\Gamma_R^{-1}\big)_{lm,l'm'}X_{l'm'}.
	\end{align}
where
	\begin{align}\label{eq:dirty}
		X_{lm}=\sum_{f,t}\gamma_{lm}^* (t;f)\frac{H(f)}{P_I(t;f) P_J(t;f)}C_{IJ}(t;f),
	\end{align}
	\begin{align}\label{eq:fisher}
		\Gamma_{lm,l'm'} = \sum_{f,t} \gamma_{lm}^*(t;f) \frac{H^2(f)}{P_I(t;f) P_J(t;f)}\gamma_{l'm'}(t;f).
	\end{align}
	
Equation \ref{eq:dirty} defines the ``dirty map'' of the SGWB convolved with the response antenna pattern of the $IJ$ detector pair. The covariance matrix of the dirty map is given in Eq. \ref{eq:fisher}. Since Eq. \ref{eq:CleanMap} represents the deconvolution of the GW signal from the detector response, we refer to it as the ``clean map''. We can obtain the covariance matrix of this clean map by inverting the covariance matrix of the dirty map. Because of this property, we refer to the matrix defined in Eq. \ref{eq:fisher} as the Fisher information matrix \cite{sph_methods}.

Inverting the Fisher matrix, however, is nontrivial, as gaps in the detector network's sensitivity across the sky cause singular values within the Fisher matrix. We therefore must regularize the Fisher matrix prior to inversion. In Eq. \ref{eq:CleanMap} the subscript $R$ denotes that the inverted Fisher matrix has been regularized. We utilize the singular value decomposition (SVD) method for this regularization, choosing which eigenmodes to discard and which to retain prior to inversion. In Section \ref{sec:point source} we consider several options for implementing the SVD regularization.

We note that the spherical harmonic expansion defined in Eq. \ref{eq:plm_sum} is cut off at some $\ell_{max}$. The choice of $\ell_{max}$ is an assumption of the search. In past searches, this value was chosen based on a {\it diffraction limit} argument in which a diffraction-limited spot size $\theta$ for a particular baseline is defined using the distance $d$ between detectors in a baseline pair and the most sensitive frequency $f_\alpha$ of the baseline (which  depends on the assumed spectral model):

\begin{align}\label{eq:lmax}
    \ell_{\rm max}=\frac{\pi}{\theta},
\end{align}
where for the LIGO-Hanford and LIGO-Livingston detectors
	\begin{align}\label{eq:diffraction}
	\theta=\frac{c}{2df_\alpha}\approx \frac{50 \rm{Hz}}{f_\alpha}.
	\end{align}

Whether or not this argument is appropriate or sufficient for making optimal GW signal detections is a question we address in Sections \ref{sec:point source} and \ref{sec:multi}.

\section{Source of Angular Resolution Limitation}

% AAM: feel free to comment out the next line
% if you don't want to see this section

%{\color{red} This section is just a draft and is poorly written, missing references, is incomplete, etc.}

%Traditionally, searches for the anisotropic gravitational-wave background have stated that the angular resolution is limited by the so-called \emph{diffraction limit}. For a monochromatic source of frequency $f$ and with two detectors separated by a distance $d$, the diffraction limited resolution $\Delta \theta$ is given by
%\begin{equation}
%    \Delta \theta = \frac{c}{2 f d}
%\end{equation}
The diffraction limit relation in Eq. \ref{eq:diffraction} gives a reasonable order-of-magnitude estimate for the angular resolution and captures the intuition that a larger baseline should improve the angular resolution \cite{mukherjee2019}. However, there are important shortcomings of this formula as usually presented. First, the word ``diffraction'' gives a misleading picture of the physics, since the angular resolution limitation actually comes from phase coherence between sources at different angular separations (as discussed below in more detail). Second, gravitational-wave sources are typically not monochromatic, and it is not clear a-priori how to transfer this diffraction limit to broadband sources. Therefore a more detailed investigation is needed to pin down the angular resolution of a given GW detector network.

%For CBC signals, it is well-known that the sky localization of the source scales inversely as (what power??) of the signal-to-noise ratio.
%(quote number for 170817?)
An important subtlety is that for deterministic signal models, such as transient CBC signals or isolated rotating neutron stars, the angular resolution of the recovered signal is evidently not limited by the diffraction limit. The CBC signal model \emph{assumes} (quite reasonably) that there is one source explaining the gravitational-wave signal. This allows triangulation of the arrival times of the signal, along with the antenna patterns, to pin down the sky location of a source to (in principle) an arbitrary good localization---currently the best localization is of order 10 square degrees for the highest SNR events seen to date. On the other hand, the anisotropic SGWB search is making an unmodeled determination of the GW sky. This generality comes with a loss in angular resolution, but does not require an assumption of a single point-like source, which is not appropriate for large scale structures. In other words, 
%Effectively the anisotropic stochastic search is looking to create an image of the GW sky, much like the Einstein Telescope is looking to create an image of a black hole. Note that a CBC search cannot create an image of the sky with only a single transient event -- 
a skymap produced by the CBC search is a probability distribution for the location of \emph{one} source, while the anisotropic SGWB search skymap is an image of the entire population of GW sources across the sky.

%In the radio-astronomy domain, the problem of combining the phases of different detectors to create an image is known as \emph{aperture synthesis}. 

In addition to the phase coherence, there is an additional effect due to blind spots in the detector network. This problem is a manifestation of the well-known \emph{bias-variance tradeoff.} As noted above, the Fisher matrix is usually not invertible, meaning that formally the variance is infinite. Regularizing the Fisher matrix reduces the variance, but at the cost of introducing a bias in the recovered signal. 

%Inspired by this, we briefly describe a toy model that can be used to understand the diffraction limit. In two dimensions, using hypothetical "scalar" gravitational-wave detectors with an isotropic that are rotating on the "Earth", the Fisher matrix is given by $\Gamma \sim J_n (...)$, where $n$ is the Fourier wavenumber ("circular harmonic.") We can see how the angular resolution is determined by the network geometry.

Inspired by this, we briefly describe a toy model that can be used to gain insight into the origin of angular resolution of the SGWB search. We consider a pair of detectors rotating on the boundary of a circle of diameter $L$ and observing scalar waves propagating in the same plane (e.g. two water buoys measuring the height of passing water waves). The detectors rotate with frequency $f_E$ such that $f_E \ll f$. Our detectors will have a response function $F(\theta)$, where $\theta$ denotes the polar angle in the plane. We consider two special cases: (a) an isotropic response $F(\theta)\propto 1$ and (b) a directional ``lighthouse'' response $F(\theta)\propto \delta (\theta-\theta_0)$, where $\theta_0$ is the opening angle of the detector. 

In the circular harmonic basis, we can define our overlap reduction function as (equivalent to Eq. \ref{eq:gammalm}):

 \begin{equation}\label{eq:circular_gamma}
\gamma_n(t,f)=\frac{1}{2}\int d\theta e^{in\theta}F_1(\theta,t)F_2(\theta,t)e^{2\pi ifL\hat{\Theta}\cdot \hat{n}(t)/c}, 
 \end{equation}
where $\hat{\theta}$ is the unit vector pointing in the direction of $\theta$ and $\hat{n}$ points from detector 2 to detector 1. 
In the isotropic response case, a suitable choice of coordinate axes allows us to rewrite our overlap reduction function as
 \begin{equation}\label{eq:circular_gamma_iso}
\gamma_n(t,f)=\frac{1}{2}e^{in(2\pi f_Et)}J_n(2\pi fL/c), 
 \end{equation}
where $J_n$ are the Bessel functions. In the lighthouse response case, this function reduces to 
 \begin{equation}\label{eq:circular_gamma_lighthouse}
\gamma_n(t,f)=\frac{1}{2} e^{in2\pi f_E t}e^{i2\pi fL\cos(2\pi f_E t)/c}\delta^2(\phi_1,\phi_2), 
 \end{equation}
where $\phi_1,\phi_2$ are the directions of detectors 1,2 at time $t=0$. 
The cross correlation between the two detectors is defined as in Eq. \ref{eq:crosscorr}, with its expectation value defined as 
 \begin{equation}\label{eq:crosscorr_circular}
\langle C(t,f)\rangle=\sum_n \gamma_n(t,f) \mathcal{P}_{n} 
 \end{equation}
We are now ready to analyze the angular resolution of this toy model. Following the anisotropic SGWB search formalism, we define the likelihood function
 \begin{equation}\label{eq:likelihood_circular}
\ln(\mathcal{L})\propto \sum_t \frac{|C(t,f_0)-\gamma_{n}(t,f_0)\mathcal{P}_{n}|^2}{P_0^2}, 
 \end{equation}
where for this model we will only be considering a single frequency bin and $P_0$ is the noise variance in this bin. Our maximum likelihood estimator is then given as 
 \begin{equation}\label{eq:MLE_circular}
\hat{\mathcal{P}}_{n}=(\Gamma^{-1})_{n m}X_{m},
 \end{equation}
with 
\begin{equation}\label{eq:X_circular}
X_{n}=\frac{1}{P_0^2}\sum_t \gamma^*_{n}(t,f_0)C(t,f_0)
\end{equation}
\begin{equation}\label{eq:Gamma_circular}
\Gamma_{nm}=\frac{1}{P_0^2}\sum_t \gamma^*_{n}(t,f_0)\gamma_{m}(t,f_0).
\end{equation}

We first consider the case of an isotropic response. The Fisher matrix $\Gamma_{mn}$, after summing over $t$, reduces to
\begin{equation}\label{eq:Gamma_circular_iso}
\Gamma_{mn}=\frac{1}{4 P_0^2}J_n^2(2\pi f_0 L /c) N_S \delta_{mn},
\end{equation}
where $N_S$ is the number of time segments. Since the Fisher matrix is diagonal, we see that all eigenvalues are nonzero for general $2\pi f_0 L /c$. However, $J_n(2\pi f_0 L/c)$ peaks around $2\pi f_0 L /c = n_0$, so wave numbers $n>2\pi f_0 L /c$ are suppressed. The effect of this eigenvalue suppression is quantitatively similar to the diffraction limit argument (c.f. Eq. \ref{eq:diffraction}), and it applies even for arbitrarily long observation times (and, by extension, high SNRs) since all eigenvalues scale equally with these quantities. We therefore see that for an isotropic response, a detector pair's resolution is limited to angular scales  similar to those given by the diffraction limit argument, albeit for a very different physical reason. 

Next, we consider the lighthouse detector response function and apply the same analysis. We again obtain a diagonal Fisher matrix that reduces to
\begin{equation}\label{eq:Gamma_circular_lighthouse}
\Gamma_{mn}=\frac{1}{4 P_0^2}\delta_{mn}\delta^2(\phi_1,\phi_2),
\end{equation}
for which all eigenvalues are equal implying that there is no limit to the angular resolution, i.e. the resolution can be arbitrarily improved by having a larger SNR signal. We conclude therefore that detector networks with perfectly isotropic responses have angular resolutions similar to those set by the diffraction limit argument, while perfectly directional detectors have unlimited angular resolution. 
In practice, however, the LIGO-Virgo detectors have anisotropic responses but not perfectly so. That is, they have responses somewhere between the two extreme cases considered above. Consequently, we should expect the anisotropic SGWB search angular resolution to be limited, but also better than what is predicted by the diffraction limit argument.

Unfortunately, the above analysis cannot be simply extended to GW detectors operating in three dimensions and across a wide frequency band with a colored noise power spectrum. Consequently, to assess the angular resolution of the anisotropic SGWB search, we resort to a series of simulations described in the following Sections. %Therefore, detailed investigations are needed to determine the optimal sky localization for detection and characterization of the background. Note that different applications of sky maps may lead to different conclusions about the optimal resolution.

\section{Single Point Source}\label{sec:point source}
%\textcolor{blue}{Describe injection methodology and mention code used to process the data. Must also describe metrics used for determining the recovered spot size and SNRs. Should be able to give the optimal search parameters for each spectral index. Need to discuss trade-off between localization and identification of sources.}
%For the remainder of this paper we will focus on an investigation into the limits of the angular resolution of the SHD search. 

As noted above, the diffraction-limit argument in Eq. \ref{eq:diffraction} may give a reasonable order-of-magnitude estimate of the angular resolution for the anisotropic SGWB search. However, this argument is inaccurate for multiple reasons. First, the angular resolution of this search is not limited by a diffraction process of GWs interacting with the detectors but rather by the phase coherence between detectors. As discussed in the previous Section, the directionality of the detector response plays an important role in the angular resolution, which is not accounted for in the diffraction limit argument. Second, when considering a network of multiple baselines, each with different separation distances and peak sensitivities, Eq. \ref{eq:diffraction} cannot be directly applied. Third, the question of regularization of the Fisher matrix (which is also related to the number of detector baselines in the network and to their relative sensitivities) also directly impacts the angular resolution of the search.

To assess the angular resolution limitations, we resort to a series of simulations and we cast the problem in terms of finding the search parameters that give optimal results when performing a spherical harmonics decomposition (SHD) SGWB search. These parameters include the frequency range over which the search is performed, the highest-order spherical harmonic modes that are considered, and the method/threshold used for Fisher matrix regularization. These parameters will differ depending on the signal model, and we will address what constitutes optimal results. 

\begin{figure}[htbp]
\centering
\includegraphics[width=0.9\columnwidth]{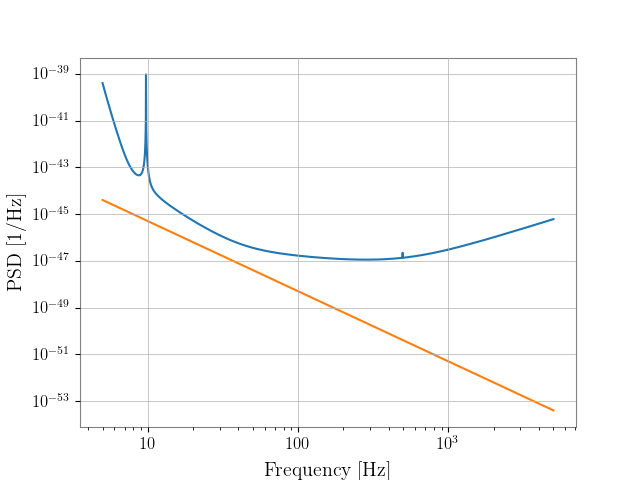}
\caption{Power spectral density (PSD) of Advanced LIGO design sensitivity (blue, upper) is shown in comparison with the strain power spectrum of the simulated point source with $\alpha=2/3$ (orange, lower). }
\label{fig:injection_psd}
\end{figure}

%We will judge a given set of search parameters based on the SHD search's ability to recover injected gravitational-wave signals. 
We begin by considering a broadband point-source GW simulation. While the SHD analysis is particularly well-suited for the recovery of spatially-extended sources, a point-source recovery is still possible with the SHD analysis, and the angular resolution of the recovery can be quantified in terms of $\ell_{max}$. In this analysis, we use the same data processing pipeline used to perform the O3 anisotropic search \cite{o3aniso} and outlined in Section \ref{sec:shd}. The point-source simulation assumes a power-law spectral shape as defined in Eq. \ref{eq:PS_spectrum}:
\begin{equation}\label{eq:PS_spectrum}
AH(f) =A\left(\frac{f}{f_{\rm ref}}\right)^{\alpha -3}, 
\end{equation}
where we choose $\alpha=2/3$ and vary the amplitude $A$. Fig. \ref{fig:injection_psd} illustrates the strain power spectrum of this simulation for  $A=5\times 10^{-49}$, alongside the power spectral density (PSD) of Advanced LIGO's design sensitivity. The single point-source was simulated at $0^{\degree}$ declination and $12^h$ right ascension.
Each simulation was 24 hours long, and for each detector the SGWB was added to the detector noise corresponding to the Advanced LIGO or Advanced Virgo design sensitivities (which are not taken to be the same). Recoveries were performed using a frequency range of 20-500 Hz and spectral index of $\alpha = 2/3$. 
%\textcolor{blue}{Summary of post-processing and calculation of SNR.}
%\begin{figure}[H]
%\centering
%\includegraphics[width=0.9\columnwidth]{SNR_map_a2_L4_H1%L1_PS5e49_mid_fk66_24hr.png}
%\caption{An SNR map showing the recovery of an injected %point-source of amplitude $\num{5e-49}$ using the HL %baseline, $l_{max}=4$, $\alpha=2/3$, and 20-500Hz %frequency range.}
%\label{fig:snr_map}
%\end{figure}

%We will first discuss the optimization of our injection recovery in terms of how well we can localize the recovered source using the Hanford-Livingston (HL) baseline. We 

\begin{figure*}[htbp]
\centering
\includegraphics[width=0.9\textwidth]{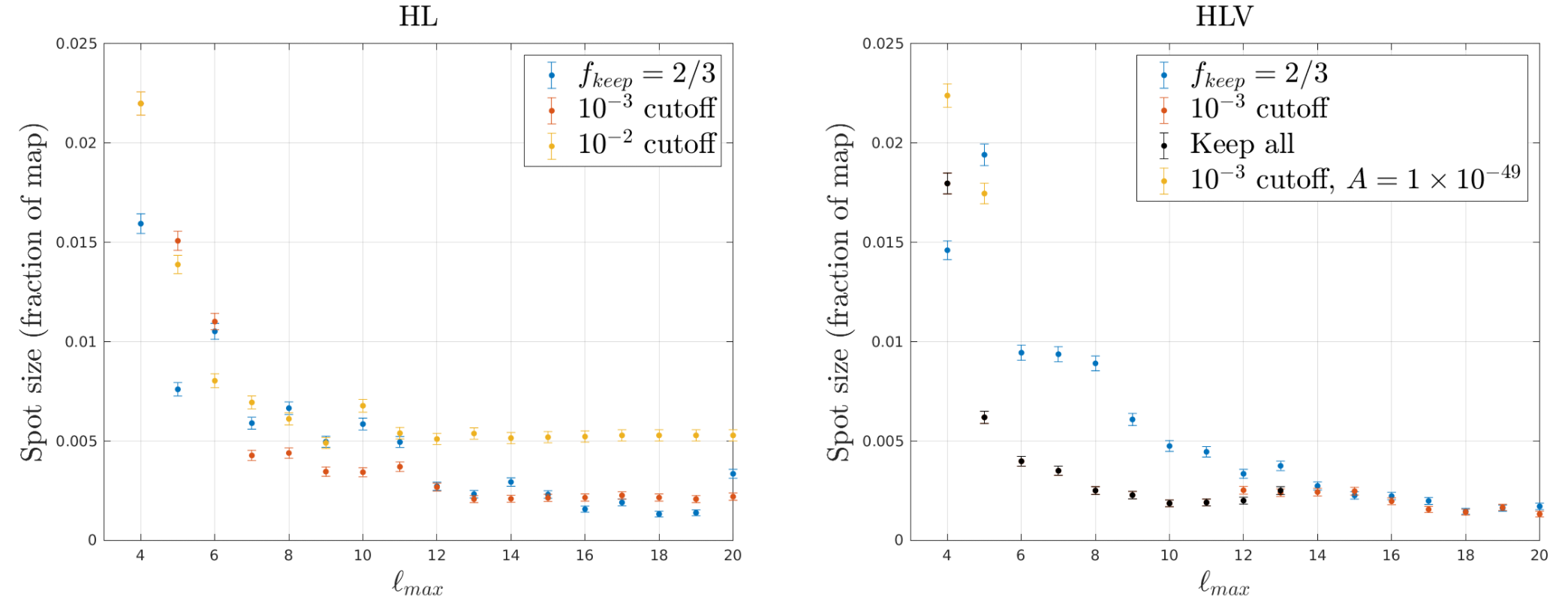}
\caption{Size of recovered point-source signal as a function of $\ell_{max}$ for multiple regularization thresholds for the Fisher matrix. The left plot shows results for the Hanford-Livingston baseline, and the right plot shows results from the HLV network. For the HLV case, we also consider the unregularized Fisher matrix (keep all). Data points are excluded if the point-source was not recovered.}
\label{fig:spot_size_HL_HLV}
\end{figure*}

To quantify the localization of the recovered point source, we count the number of pixels in the recovered SNR skymap whose SNR is larger than 75\% of the map's peak SNR value. The spot size defined in this way is shown in Fig. \ref{fig:spot_size_HL_HLV} (left) as a function of  $\ell_{max}$. As $\ell_{max}$ is increased, the spot size is reduced and the resolution of the recovered map is improved. For comparison, for $\alpha=2/3$ the most sensitive frequency is about 65 Hz, resulting in the diffraction limit in Eq. \ref{eq:diffraction} of $\theta \approx 45^{\circ}$ and the diffraction limit spot size of about 4\% of the sky. We can also consider the fact that we make statistically significant recoveries of the simulated source at an $\ell_{max}$ value of 20, and the diffraction-limited spot size corresponding to this $\ell_{max}$ value is $\theta \approx 9^{\circ}$, which represents about 0.2\% of the sky. It is evident from Fig. \ref{fig:spot_size_HL_HLV} that the anisotropic SGWB search can substantially surpass the diffraction limit, as the smallest spot size measurements we obtain are about 0.2\% of the sky map.
%The overall trend we can see is that high $\ell_{max}$ values are optimal for localization of features in the GW skymap. 

Figure \ref{fig:spot_size_HL_HLV} (left) also compares different Fisher matrix regularization methods. The first regularization threshold we apply is to keep two thirds of all eigenvalues of the Fisher matrix ($f_{keep}=2/3$). This is the method that has been in use for all previous SHD searches for anisotropic SGWB \cite{o3aniso,Abbott:O2_aniso,o1_directional}. 
%\VUK{Add references for O1 and O2 directional papers}
The second regularization threshold we apply removes the eigenvalues that are smaller that $10^{-3}$ times the largest eigenvalue \footnote{This is often stated as the condition number of an eigenvalue being smaller than $10^{-3}$}. A comparison between these two methods is shown in Fig. \ref{fig:spot_size_HL_HLV} (left)---while the two regularization methods yield similar results, the $10^{-3}$ threshold performs slightly better. For the $10^{-3}$ threshold, a high $\ell_{max}$ makes little difference to the spot size since most new eigenvalues introduced by raising $\ell_{max}$ are small and therefore removed by the regularization threshold. This behavior can be seen in the relatively constant spot size measurements for high $\ell_{max}$ values in Fig. \ref{fig:spot_size_HL_HLV} (left). We also include spot size measurements that result from using a more aggressive regularization scheme in which we apply a $10^{-2}$ threshold. This threshold results in more eigenmodes being discarded during the Fisher matrix regularization, giving larger spot size measurements. %\VUK{we should revisit this statement once you impose a cut in significance. It is possible that the flatness of this curve at high lmax is due to failing to recover the right spot}

%When we use the $f_{keep}=2/3$ regularization threshold, it is possible that high values of $\ell_{max}$ begin to include modes to which we are not sensitive, preventing us from recovering the source at all and giving spot size measurements that are no longer meaningful. However, for the range of $ell_{max}$ For the $10^{-3}$ regularization threshold, a high $l_{max}$ makes little difference to the spot size since most new eigenvalues introduced by raising $l_{max}$ are small and therefore removed by the regularization threshold. 

%\begin{figure}[H]
%\centering
%\includegraphics[width=0.9\columnwidth]{spot_size_5e49.png}
%\caption{Size of recovered point-source signal as a function of $l_{max}$ for %two different regularization thresholds.}
%\label{fig:spot_size}
%\end{figure}

We also consider a modified version of this search in which we include the Virgo detector, resulting in a network of three detector baselines. We denote this network as HLV. A single baseline will not be sensitive to all sky directions, and these gaps in sensitivity manifest as singular Fisher matrix eigenvalues. When more detector baselines are included in a network, it is possible for the Fisher matrix to become naturally regularized as the network gains sensitivity to more sky directions. A different Fisher matrix regularization strategy may therefore be better suited for multi-baseline detector networks. The size of the recovered point-source with this multi-baseline network is shown in Fig.~\ref{fig:spot_size_HL_HLV} (right). In addition to the above two regularization thresholds, we also consider using the entire (unregularized) Fisher matrix, i.e. keeping all of its eigenvalues.
%\begin{figure}[H]
%\centering
%\includegraphics[width=0.9\columnwidth]{spot_size_HLV_power_compare_20_present%ation.png}
%\caption{Size of recovered point-source signal as a function of $l_{max}$ for %three different regularization thresholds using the HLV network.}
%\label{fig:spot_size_HLV}
%\end{figure}

%\begin{figure}[H]
%\centering
%\includegraphics[width=0.9\columnwidth]{fisher_lmax_4.png}
%\caption{Fisher matrix conditioning for the H1L1 and HLV networks at %$l_{max}=4$.}
%\label{fig:fisher_lmax_4}
%\end{figure}

\begin{figure*}[htbp]
  %\begin{tabular}{c}
	%% SHD (O1+O2)HL + (O3A)HLV SNR maps  	
  \includegraphics[width=\textwidth]{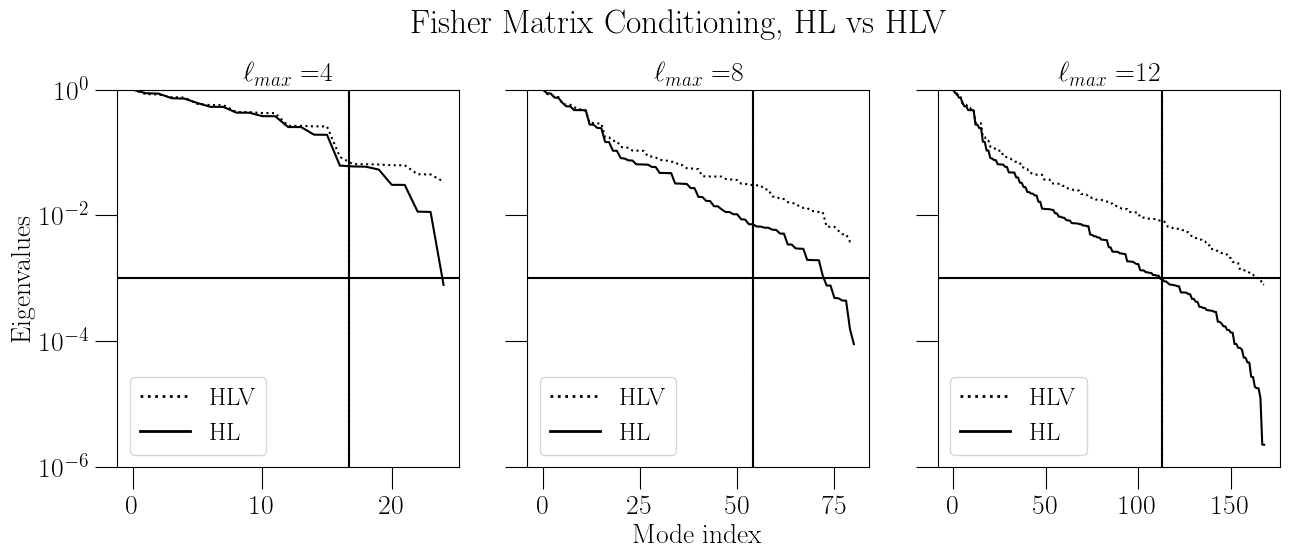}
	%% SHD Upper Limit sky maps  	
  %\end{tabular}
  \caption{%% modified text In the top row are the spherical harmonics SNR
  Fisher matrix conditioning plots for $\ell_{max}=4,8,$ and $12$ comparing the HL and HLV detector networks using simulated data. Vertical lines mark the largest 2/3 of eigenvalues. Horizontal lines mark the $10^{-3}$ threshold.}

\label{fig:fisher_condition_HLV}  
\end{figure*}

For $\ell_{max}<12$, using the $10^{-3}$ threshold is equivalent to keeping all eigenvalues. The $10^{-3}$ threshold generally gives the most localized recoveries of the three regularization methods used. For $\ell_{max}\geq12$, the $10^{-3}$ threshold performs similarly to the $f_{keep}=2/3$ method. For $\ell_{max}>13$, the recovered peak SNR values are not statistically significant when we keep all eigenvalues, so we conclude that the source is not properly recovered in this case, and we do not report spot size values. We can conclude therefore that some regularization is still necessary for this multi-baseline network for $\ell_{max}>13$. If we decrease the amplitude of our simulated source from $A=5\times 10^{-49}$ to $A=1\times 10^{-49}$, we only make statistically significant recoveries of the source for $\ell_{max}<6$, and when we do recover this lower-amplitude source, its spot size measurements are larger than those of the higher-amplitude source as seen in Fig. \ref{fig:spot_size_HL_HLV} (right). This is consistent with our expectation that larger SNR allows recovery with better angular resolution. %sporadically when we choose to keep all eigenvalues. These scattered spot size measurements correspond to instances in which the injected source is not being accurately recovered.  \VUK{have to revisit this paragraph after you impose the significance cut.}

Figure \ref{fig:fisher_condition_HLV} directly compares Fisher matrix conditioning in the HL and HLV cases. We see that for a low $\ell_{max}$ value like $\ell_{max}=4$, there is little difference in the Fisher matrix eigenvalues between the HL and HLV networks if we only consider eigenvalues kept using the $f_{keep}=2/3$ threshold. If we consider all the eigenvalues, the smallest eigenvalue in the HLV Fisher matrix is an order of magnitude larger than that of the HL Fisher matrix.
%\VUK{I don't see this, the HLV and HL Curves are about the same at the 2/3 cutoff}
Using higher values of $\ell_{max}$ shows a widening in the differences between the HL and HLV Fisher matrix conditioning, with the HLV network having larger eigenvalues even among those that are accepted by the $f_{keep}=2/3$ regularization method. This improvement in conditioning is likely responsible for the slightly smaller spot size measurements for HLV (as compared to HL) shown in Fig. \ref{fig:spot_size_HL_HLV}. 

We next examine the impact of the chosen frequency band on the conditioning of the Fisher matrix. For the spectral index $\alpha=2/3$, the most sensitive frequency is $\sim 65$ Hz \cite{abbott2017directional}. Further, 99\% of the HL baseline's sensitivity lies within the 20-120 Hz band for the search for an isotropic SGWB with $\alpha=2/3$ \cite{isotropic2020}. We therefore start with the 20-100 Hz frequency band that captures most of the sensitivity of the SHD search. We compare the conditioning of the Fisher matrix in this band with the conditioning in wider bands (20-200 Hz and 20-500 Hz)---these wider bands include higher frequencies and therefore may be sensitive to smaller angular scales on the sky.
%We can also consider using a smaller frequency range that primarily covers only those frequencies to which our detector network is most sensitive.  If we choose to run our SHD search using a frequency range of 20-100Hz as opposed to 20-500Hz, we can see whether this choice affects our search by again looking at the conditioning of the Fisher matrix (Fig.\ref{fig:fisher_condition_freq}).

\begin{figure*}[htbp]
  %\begin{tabular}{c}
	%% SHD (O1+O2)HL + (O3A)HLV SNR maps  	
  \includegraphics[width=\textwidth]{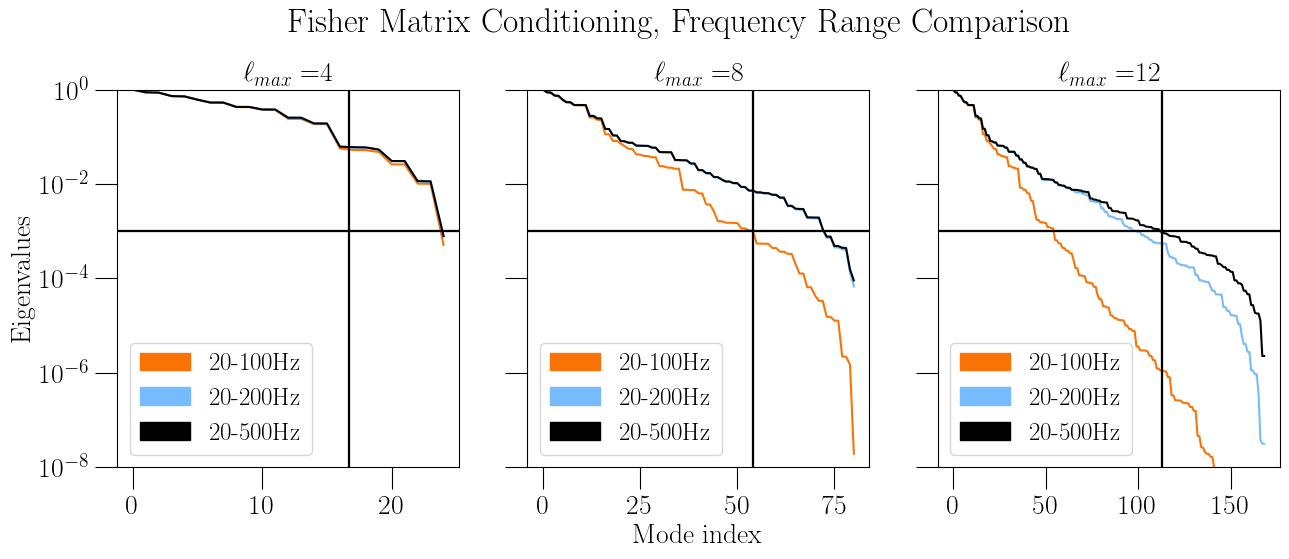}
	%% SHD Upper Limit sky maps  	
  %\end{tabular}
  \caption{%% modified text In the top row are the spherical harmonics SNR
  Fisher matrix conditioning is shown for $\ell_{max}=4,8,$ and $12$, comparing several different frequency bands for the HL baseline. Vertical lines mark the lowest 1/3 of eigenvalues and horizontal lines mark the $10^{-3}$ threshold.}

\label{fig:fisher_condition_freq}  
\end{figure*}

Figure \ref{fig:fisher_condition_freq} shows that as the $\ell_{max}$ value increases, the regularization of the Fisher matrix worsens drastically for the 20-100 Hz band. Extending the frequency band to 20-200 Hz gives a marked improvement, but the conditioning is the best when using the widest frequency band of 20-500 Hz. This difference in conditioning implies that including higher frequencies allows the detector network to resolve smaller-scale structures in the skymap, thereby increasing the angular resolution of the recovery. 

\begin{figure}[htbp]
\centering
\includegraphics[width=0.9\columnwidth]{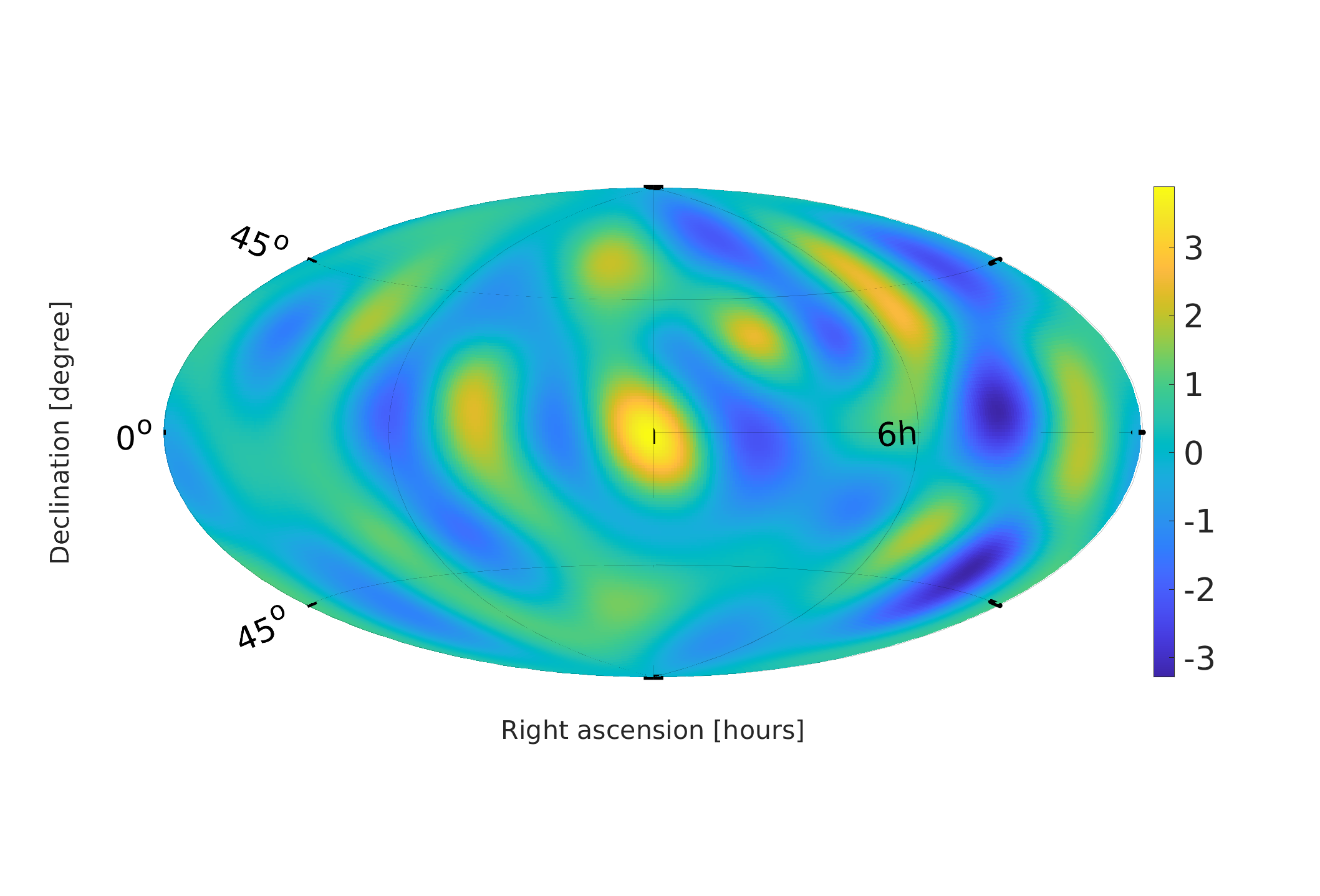}
\caption{An example SNR map is shown of a simulated point-source with amplitude $A=10^{-49}$ and location $0^{\degree}$ declination, $12^h$ right ascension, recovered using the HL baseline,  $f_{keep}=2/3$, and $\ell_{max}=8$.}
\label{fig:snr_l8}
\end{figure}

\begin{figure}[htbp]
\centering
\includegraphics[width=0.9\columnwidth]{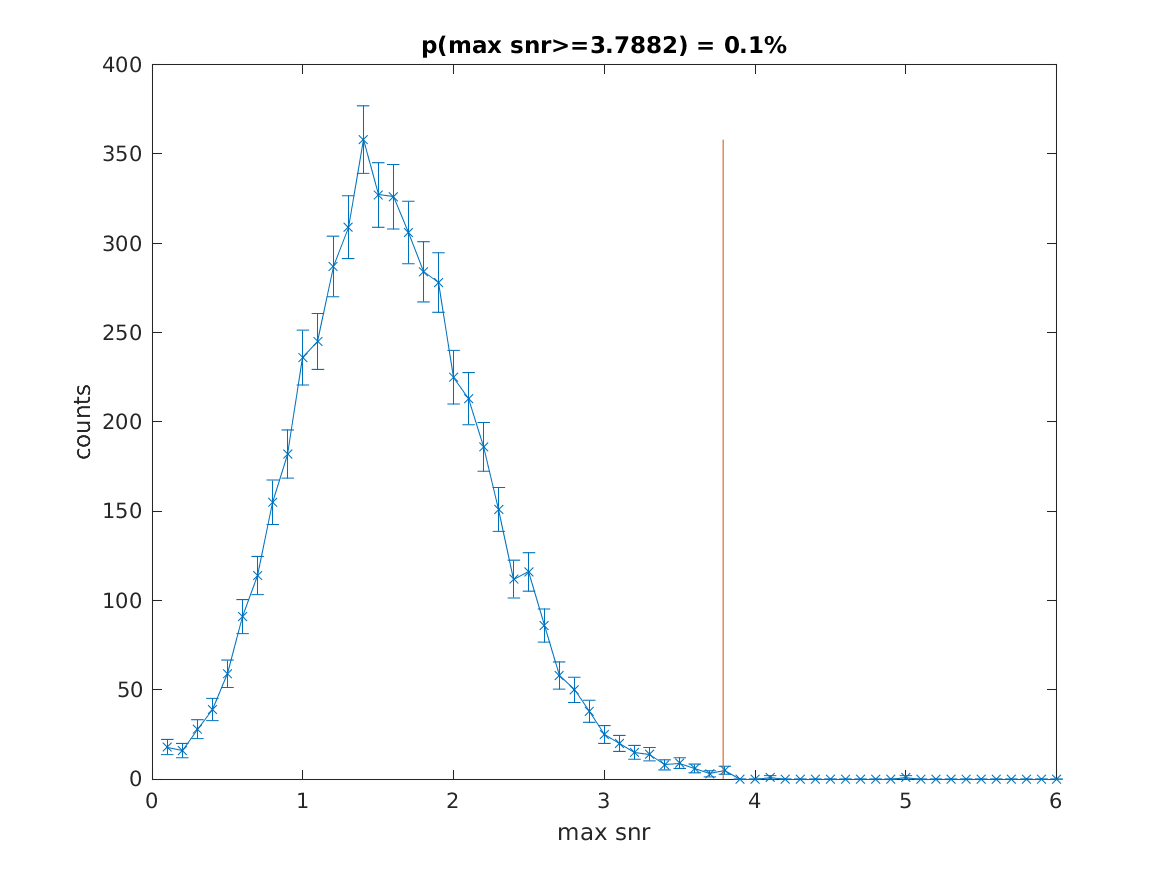}
\caption{Peak SNR from the map in Fig. \ref{fig:snr_l8} is marked by a vertical line in comparison with the histogram of the peak SNRs obtained from purely noise simulated skymaps.}
\label{fig:sig_hist}
\end{figure}

Finally, we examine our ability to detect a point source rather than our ability to localize it. We quantify the significance of a GW source recovery in a manner similar to that used in previous directional searches \cite{o3aniso}. We draw from a multivariate Gaussian distribution with zero mean and covariance given by the inverted Fisher matrix, resulting in a set of $P_{\ell m}$ values corresponding to a noise skymap. %We apply random scaling factors to the standard deviations $\sigma_{lm}$ of the clean map estimators $\mathcal{P}_{lm}$ in order to produce many randomized realizations of the clean map which are consistent with noise. %This method involves randomizing the eigenvalues of the Fisher matrix many times to produce many different randomized realizations of the clean map that are based on the the original data. 
The procedure is repeated 5000 times and an SNR map is made for each such random realization of noise. In each SNR map we define a region centered around the location of the simulated source, extending $20^{\degree}$ above and below the simulated point source and $2^h$ on either side. We then find the mean pixel value among the top 5\% of the pixels within this region and bin this value into a histogram. We make the same measurement using a skymap containing the simulated point source signal, like the one shown in Fig. \ref{fig:snr_l8} for $\ell_{max}=8$. We attribute a (significance) p-value to the peak SNR in this map by comparing it to the noise peak-SNR histogram. An example is shown in Fig. \ref{fig:sig_hist}, where the vertical line denotes the peak SNR measurement of the map in Fig. \ref{fig:snr_l8}.

\begin{figure}[htbp]
\centering
\includegraphics[width=0.9\columnwidth]{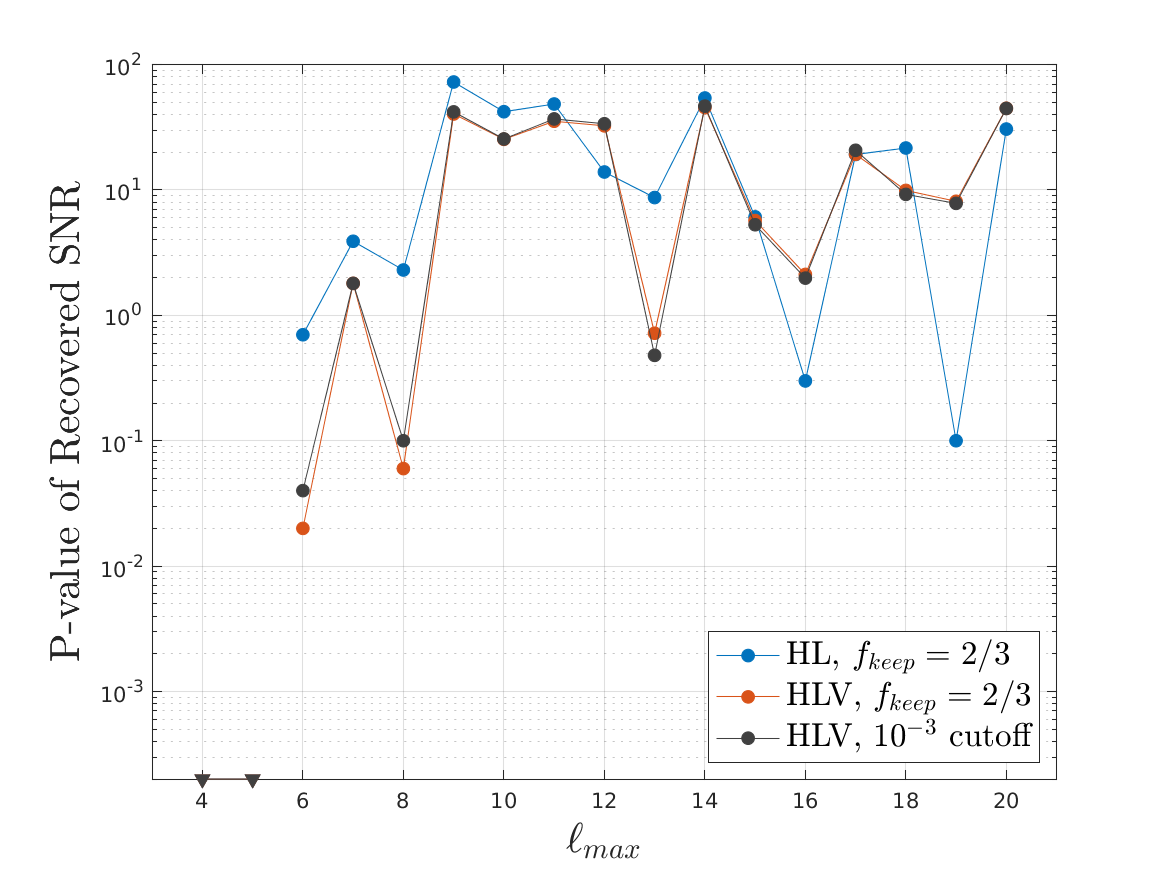}
\caption{We show the significance (p-value) of the recovered simulated peak SNR vs $\ell_{max}$ using HL baseline, $f_{keep}=2/3$, and point-source amplitude $A=10^{-49}$. p-values equal to zero are denoted with downward-pointing arrows at the bottom of the plot. We also show results for the HLV network, for two different regularization methods. We see that using higher $\ell_{max}$ gives recovered peak SNRs with lower significance (larger p-values).}
\label{fig:pvalue_vs_lmax}
\end{figure}

We find that increasing $\ell_{max}$ causes an increase in the p-value associated with its peak SNR and therefore a decrease in the ability to detect the simulated source. These results are shown in Fig.~\ref{fig:pvalue_vs_lmax}. For each additional order of spherical harmonics modes we include in our analysis by increasing $\ell_{max}$, we increase the number of model parameters used to fit to the data, which ultimately results in a decrease in the SNR of the recovered source. We therefore find that low $\ell_{max}$ values lend themselves to optimal point source signal detection, while high $\ell_{max}$ values may be better suited for optimal localization of the point source signal, depending on the signal strength. In other words, search parameters that are optimal for detecting a GW source may differ from those that are optimal for localizing it.

%\textcolor{blue}{talk about over-fitting? explain why high $\ell_{max}$ values give lower SNRs}

%\begin{figure}[H]
%\centering
%\includegraphics[width=0.9\columnwidth]{fisher_lmax_4_freq.png}
%\caption{Fisher matrix conditioning for the two different frequency %ranges at $l_{max}=4$.}
%\label{fig:fisher_lmax_4_freq}
%\end{figure}

%\begin{figure}[H]
%\centering
%\includegraphics[width=0.9\columnwidth]{fisher_lmax_12_freq.png}
%\caption{Fisher matrix conditioning for the two different frequency %ranges at $l_{max}=12$.}
%\label{fig:fisher_lmax_12_freq}
%\end{figure}

\section{Multiple Point Sources}\label{sec:multi}
%\textcolor{blue}{Discuss whether two points can be resolved arbitrarily well given high enough amplitude, use similar spot size metric as previously}
%Finally, we will investigate the extent to which the simultaneous recovery of multiple point-sources is limited in this search. In the previous section, we investigated the angular resolution of our detector networks by seeing how well they could detect a single point-source, but 
%We would like to know what search conditions enable us to resolve between multiple sources rather than observing a single spatially extended signal.
%For a pair of stationary detectors, the recovery of multiple GW sources would not be feasible, since degeneracies in the baseline's sensitivity pattern would cause the signals to be averaged, giving a recovery result that does not correspond to a real GW signal. However, the rotation of our detectors should break any such degeneracies, allowing the recovery of multiple sources. 

Another way to characterize the angular resolution of the SHD search is by its ability to distinguish between multiple point sources. We begin by simulating two point-sources of equal magnitude with declinations of $+30^{\degree}$ and $-30^{\degree}$, and we measure the spot size of each of the two sources by the same process as in Fig.~\ref{fig:spot_size_HL_HLV},
but applied separately to the upper and lower halves of the map. 
We note that the choice to look at the two halves of each map separately requires knowledge of the sources' actual locations, which would not be available in real GW searches. However, the goal of this test it to investigate
the ability to resolve separated point-sources, and not to provide the optimal methodology for identifying multiple point sources in the skymap. 
Amplitudes of simulated point sources in this test were louder than those used previously in this paper. This choice was necessary so as to investigate whether neighboring GW signals with sufficiently high SNRs are better recovered when we use sufficiently high-order spherical harmonics modes.
%This choice was made so that the sources were still recoverable at high $\ell_{max}$ values, allowing us to investigate whether neighboring GW signals with sufficiently high SNRs are recoverable given we use sufficiently high-order spherical harmonics modes.
%
\begin{figure}[htbp]
\centering
\includegraphics[width=0.9\columnwidth]{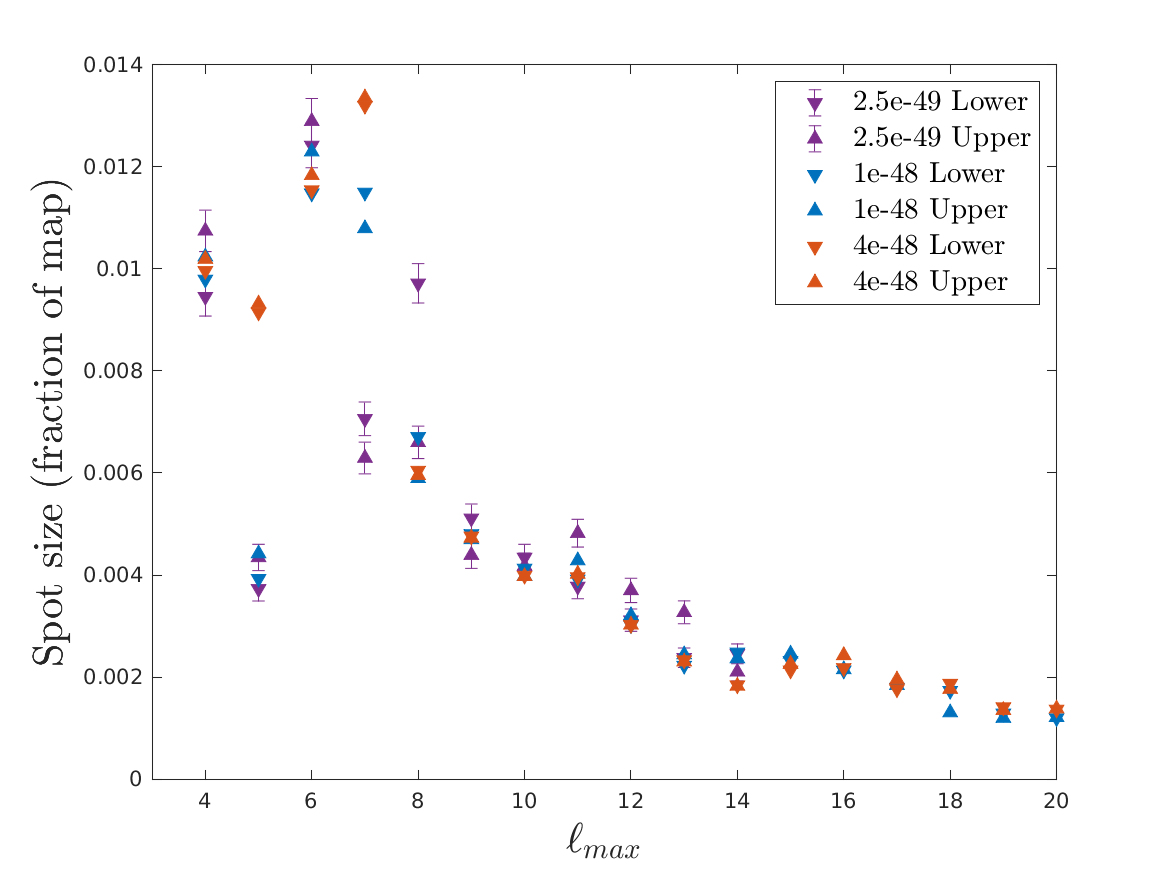}
\caption{Sizes of two recovered point-source signals as a function of $l_{max}$ are shown for multiple source amplitudes. The upper and lower point sources are denoted with upward- and downward-pointing arrows, respectively. Here we use the HL baseline. Error bars are given for one set of points to reduce clutter. Other data points have comparable errors. }
\label{fig:spot_size_multi}
\end{figure}
\begin{figure}[!htbp]
\centering
\includegraphics[width=0.9\columnwidth]{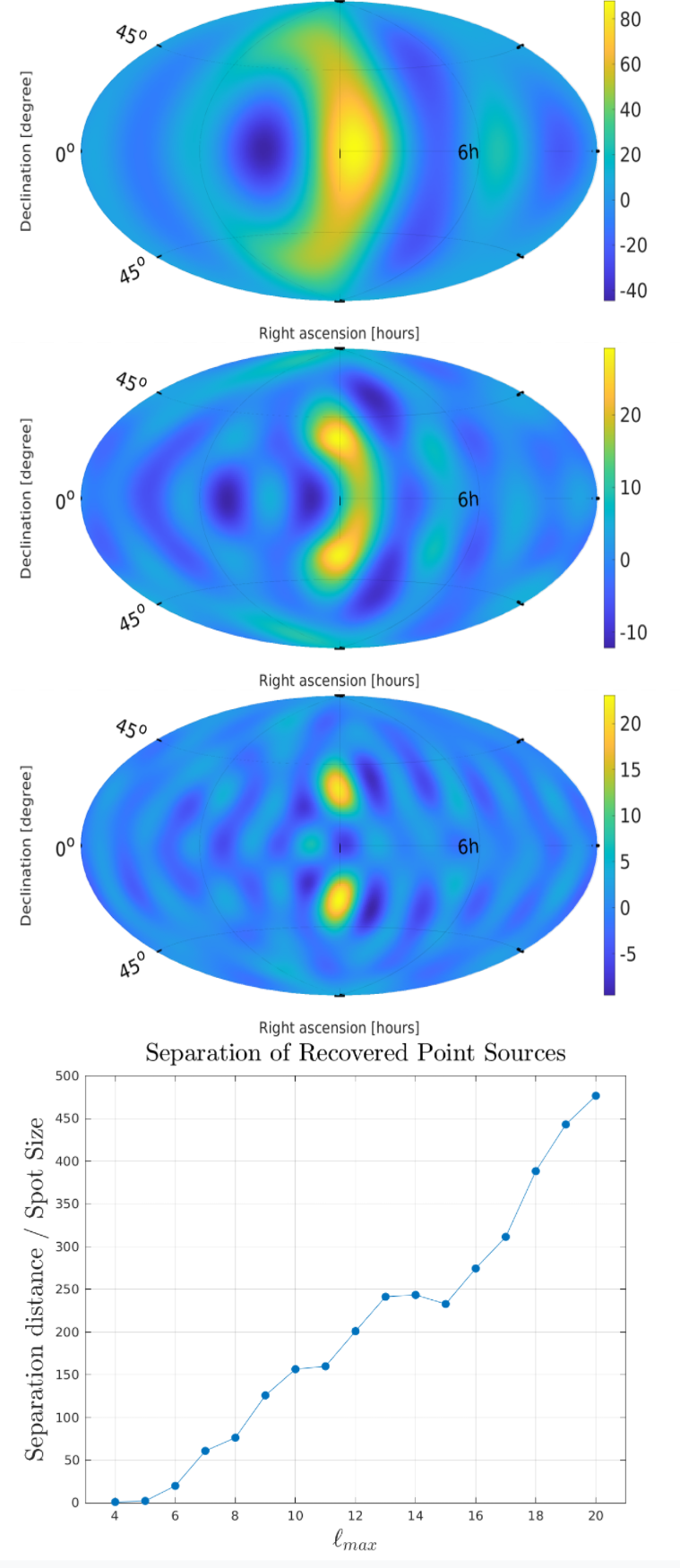}
\caption{SNR maps of two recovered point-source signals using the HL baseline are shown for point source amplitudes of $10^{-48}$ and for $l_{max}=4,8,12$ (top to bottom). In the bottom plot we also show the ratio of the angular separation to the total spot size of the upper and lower sources as we increase $l_{max}$.}%\VUK{this plot needs to be normalized so that vertical axis is 1 when the two sources are barely separated.}}
\label{fig:snr_multi}
\end{figure}

Figure \ref{fig:spot_size_multi} shows each source's spot size as a function of $\ell_{max}$.
We observe behavior similar to those of the single point-source recoveries, namely that the spot sizes of the two sources do in fact decrease as we use higher-order spherical harmonics modes. Also, we see that the upper and lower spot size measurements are very similar in most cases. However, this does not necessarily mean that we are properly resolving the two signals. 

Figure \ref{fig:snr_multi} shows the SNR maps corresponding to dual point-source simulations of amplitude $A=10^{-48}$. We see that for low $\ell_{max}$, the two sources are not resolved. Only at higher $\ell_{max}$ do we see distinct recoveries. The bottom plot of Fig.~\ref{fig:snr_multi} shows that the resolution of the two point sources improves with increasing $\ell_{max}$.

Since we used loud simulations, the sources are still recoverable at high $\ell_{max}$ despite the decreasing SNR behavior discussed in the previous Section. If we were to use sources with lower SNRs, they would only be statistically distinct from noise if we use low spherical harmonics modes, which in turn reduces the angular resolution of the map, preventing the sources from being resolved. Therefore we find that in order to recover multiple GW point sources, they must be sufficiently loud so as to support analysis with sufficiently high-order spherical harmonics components. This result also reinforces the notion that low $\ell_{max}$ should be considered when initially performing a search in order to detect the presence of a GW source, before attempting analysis with  higher order modes to optimize localization.

We note that the recovered spot sizes shown in Fig. \ref{fig:spot_size_multi} are even smaller (by about $2\times$ in area) than those shown for single point-source simulations in Fig. \ref{fig:spot_size_HL_HLV}. This is a consequence of the higher amplitude, $A=10^{-48}$, used in the simulations shown in Fig. \ref{fig:spot_size_multi}. In fact this amplitude has already been excluded by past SGWB searches with Advanced LIGO and Advanced Virgo data.

\section{Extended Sources}\label{sec:extended}
%Since the SHD search accounts for possible correlations between pixels, it is particularly suited for identifying features on the GW skymap that are spatially extended. While up until now we have considered only GW point-sources as a simple starting point in our angular resolution investigation, our ultimate goal will be to describe our search's angular resolution as it pertains to extended sources. In this section we present the results of applying our previously-described methodology to the recovery of an injected extended GW source. 

\begin{figure}[htbp]
\centering
\includegraphics[width=0.9\columnwidth]{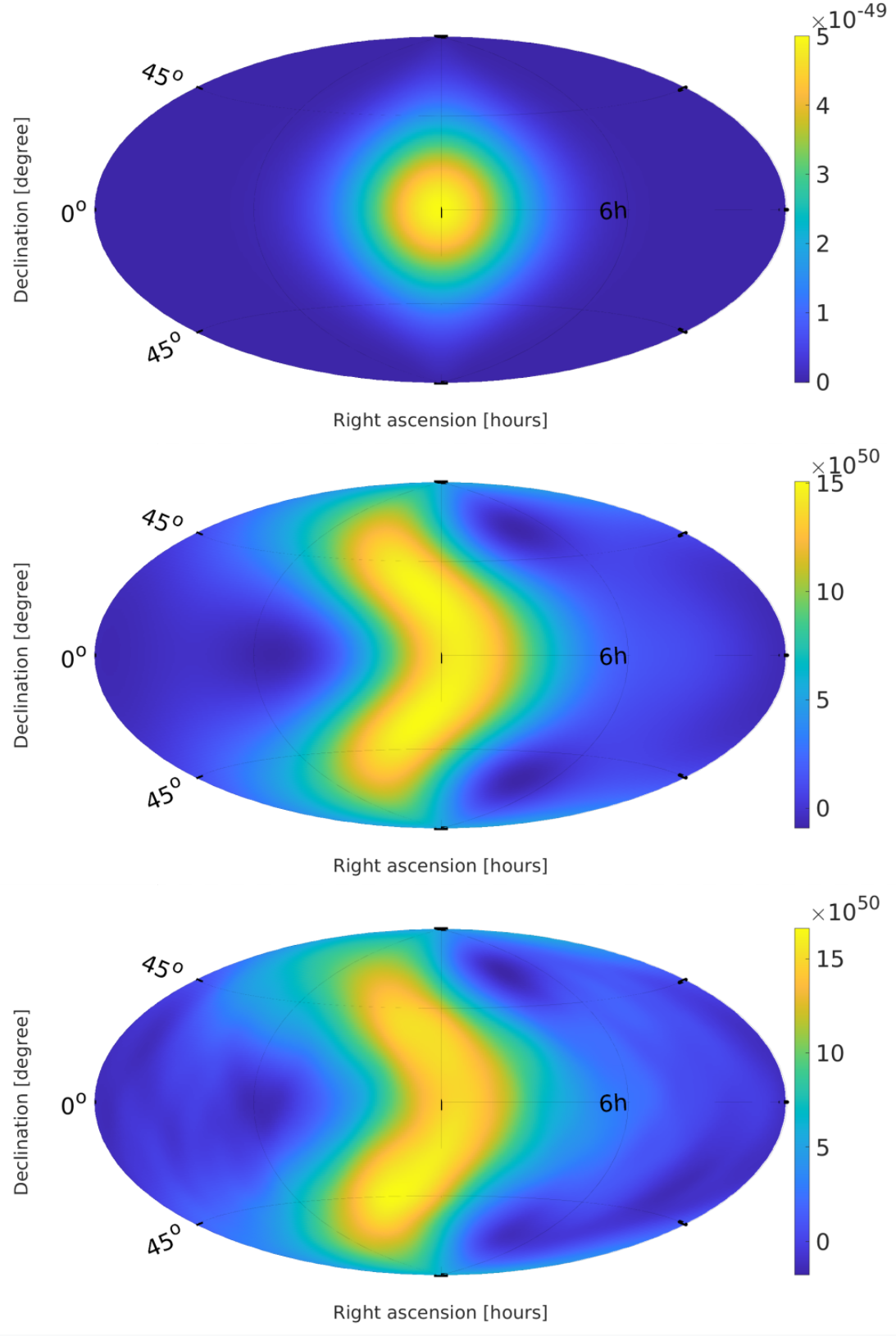}
\caption{From top to bottom, clean map of 2-D Gaussian simulation, its corresponding dirty map, and the sum of this dirty map with the noise-only dirty map.}
\label{fig:extended_maps}
\end{figure}

We extend the methodology presented in previous Sections to recovery of spatially extended sources. Extended sources come in many varieties, and we do not attempt an exhaustive study here. Instead, we chose to study an extended source in the shape of a two-dimensional Gaussian distribution centered at $0^{\degree}$ declination and $12^h$ right ascension with amplitude $5 \times 10^{-49}$ and standard deviations of $30^{\degree}$ in each direction. We first create a clean map of this signal in the HEALPix basis \cite{HEALPix}. We then perform the SHD search on noise-only data to produce a Fisher matrix and dirty map corresponding to only noise. Multiplying this noise-only Fisher matrix by the clean signal map gives the dirty map corresponding to the extended source (c.f. Eq. \ref{eq:CleanMap}). This dirty map is added to the noise-only dirty map to give a final dirty map containing both noise and the simulated source.  Fig. \ref{fig:extended_maps} shows the clean signal map, the dirty signal map, and the final combined dirty map.

\begin{figure}[H]
\centering
\includegraphics[width=0.9\columnwidth]{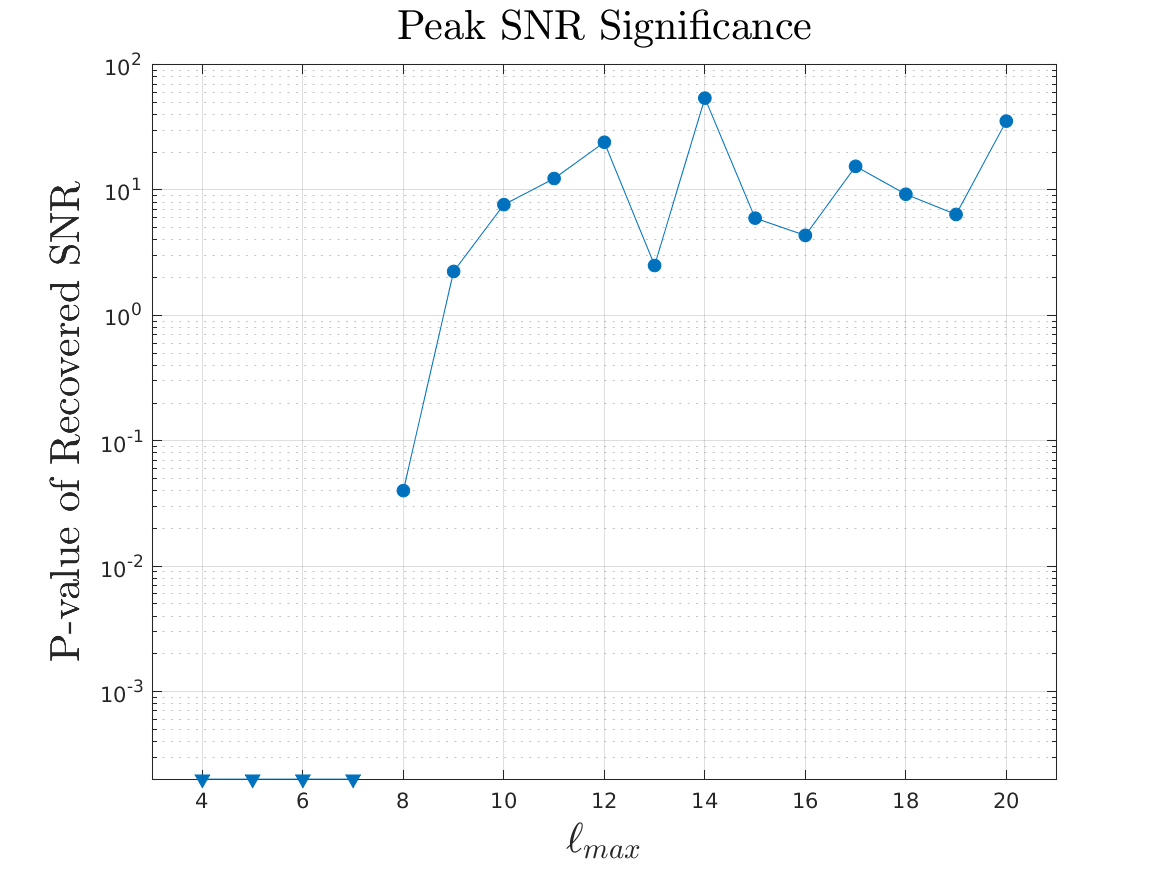}
\caption{Significance of the recovered extended source as a function of $\ell_{max}$, using the HL baseline and Gaussian source of amplitude of $5\times10^{-49}$.}
\label{fig:extended_sign}
\end{figure}

\begin{figure}[H]
\centering
\includegraphics[width=0.9\columnwidth]{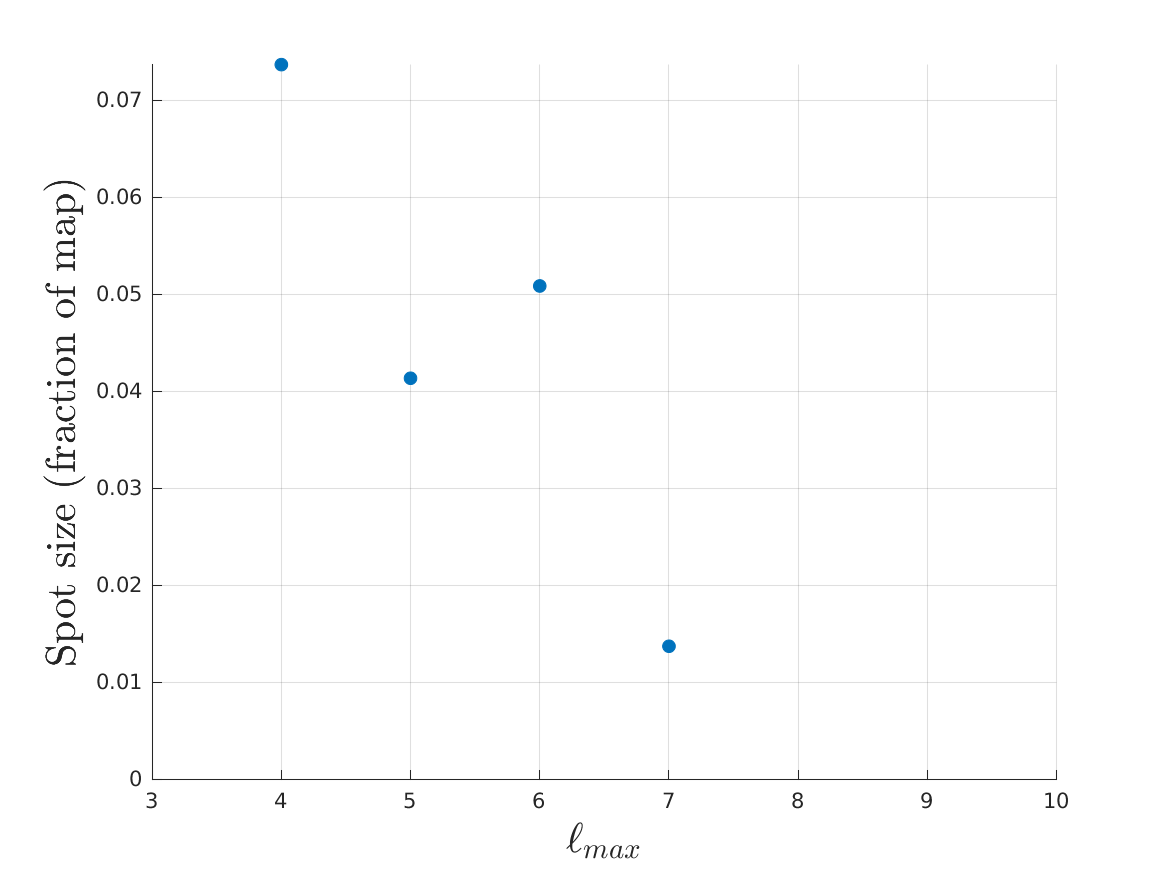}
\caption{Size of the recovered extended source as a function of $\ell_{max}$, using the HL baseline and Gaussian source of amplitude of $5\times10^{-49}$.}%\VUK{This plot should be cut off at lmax=7}
\label{fig:extended_size}
\end{figure}

\begin{figure}
\centering
\includegraphics[width=0.9\columnwidth]{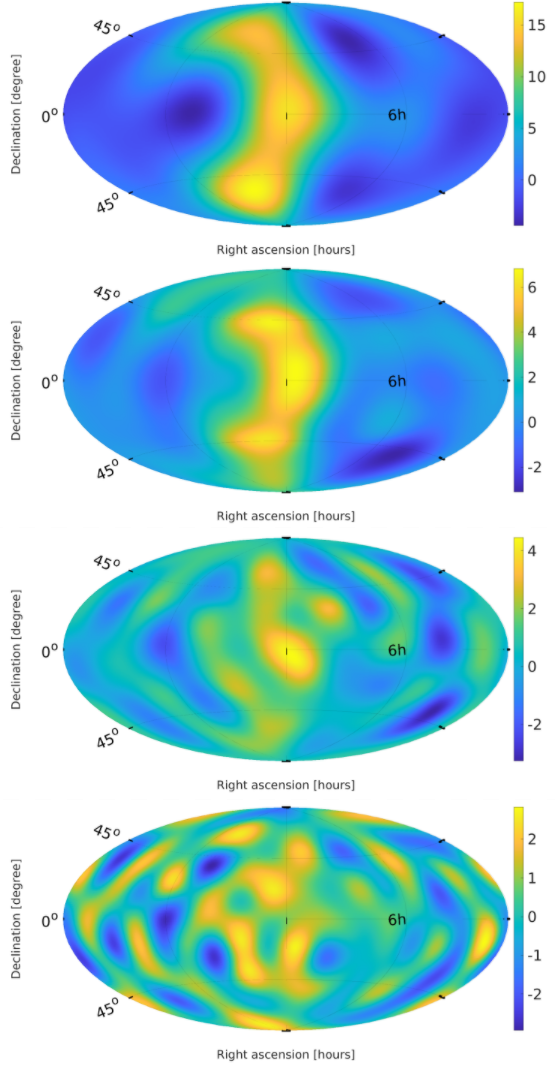}
\caption{SNR maps of the recovered extended source using the HL baseline and Gaussian source of amplitude of $5\times10^{-49}$. From top to bottom, $\ell_{max}=4,6,8,10$.}
\label{fig:extended_snr}
\end{figure}

The final combined dirty map and the Fisher matrix are then used to perform the same analysis described in Sec. \ref{sec:point source}. Namely, we obtain spot-size and significance measurements using the same methodology as in the point-source case, using the HL baseline and $f_{keep}=2/3$. Figure \ref{fig:extended_sign} shows the significance of the recovered signal as a function of $\ell_{max}$. As in the point-source case, we see the significance is the largest (p-values are the smallest) for low $\ell_{max}$ values. For this particular simulation, $\ell_{max}=8$ seems to be the cutoff beyond which our recoveries are not statistically significant. 
In Fig. \ref{fig:extended_size} we show the spot size  as a function of $\ell_{max}$. As in the point-source case, we observe the spot size decreasing as $\ell_{max}$ increases. The smallest size is reached for $\ell_{max} = 7$, at which point the recovered spot makes up about 1\% of the map, roughly consistent with the size of the simulated extended source. Attempting recovery for $\ell_{max}>7$ is not successful since the SNR of the simulation is not large enough to support analyses with larger numbers of free parameters. This is illustrated in Fig. \ref{fig:extended_snr}, where e.g. for $\ell_{max}=10$ we see no discernible recovery of the extended source. 

%While one may expect the spot size measurement of a recovered point-source to be arbitrarily small for an arbitrarily strong signal and high $\ell_{max}$, the spot-size measurement of an extended source may instead be expected to plateau at some finite size. 
%We do not observe this behavior in Fig. \ref{fig:extended_size}, which seems at first to demonstrate the same behavior we saw in the point-source case. However, it should be noted that for $\ell_{max}>8$ we do not recover the injected source at all, and our spot size measurements for those higher $\ell_{max}$ values are not meaningful. Our inability to recover this injected source at these higher $\ell_{max}$ values is 
%

Overall, we observe that recoveries of the simple spatially extended source follow the same patterns as recoveries of point sources. Specifically, low values of 
$\ell_{max}$ are appropriate for the first detection of an anisotropic SGWB, as they offer the largest statistical significance of the recovery. Larger values of $\ell_{max}$ should then be used in follow-up analyses to better localize the source, to the extent supported by the amplitude (SNR) of the source. Future work is needed for a more exhaustive study of spatially extended sources of different shapes and morphologies.

\section{Conclusions}\label{sec:conclusion}

Observations of GWs from individual compact binary coalescences have raised the possibility of observing the SGWB in the upcoming observation runs of terrestrial GW detectors. Observation of SGWB anisotropy promises to be an important source of information about the population of systems that give rise to the SGWB. Consequently, understanding the angular resolution of the anisotropic SGWB search and how it changes as a function of search parameters will be increasingly important in future analyses of GW data. 

In this paper we have investigated the impact of several search parameters on the angular resolution of the anistropic SGWB search. We have found that the use of a wide frequency band, such as 20-500 Hz, and of a detector network containing multiple baselines both help increase the number of directions (or spherical harmonic modes) to which our network is sensitive. Consequently, this improves the conditioning of the Fisher matrix, simplifies its inversion, and overall improves the angular resolution of the search. 

We have also demonstrated that for detecting the presence of an anisotropic SGWB, one should search using low-order spherical harmonics modes. Once the detection is made, follow-up analyses using higher-order spherical harmonic modes should be pursued to improve the angular resolution of the analysis and better localize the source. Since searching with increasingly higher-order modes reduces the significance of the recovery, the maximum number of modes (i.e. $\ell_{max}$) will be ultimately set by the amplitude (or SNR) of the GW source. In other words, the optimal localization of a GW source is achieved by using the highest number of spherical harmonics modes that still give statistically significant recoveries. Further, for a sufficiently strong SGWB it is possible to surpass the traditional ``diffraction limit'' on the angular resolution of the anisotropic SGWB search.  

Similarly, we have shown that multiple point sources can be recovered and resolved by the anisotropic SGWB search. If the sources' amplitudes are sufficiently large, the search could be conducted with a sufficiently large $\ell_{max}$ so as to resolve the point sources, while still providing statistically significant recovery. Spatially extended sources follow a similar pattern, as we have demonstrated on a simple case of a Gaussian-distributed extended source. 

Future studies should consider more complete sampling of shapes and morphologies of spatially extended sources. We also note that the results presented in this paper could be altered when a different spectral model is used (e.g. for spectral index $\alpha$ different from $2/3$), as well as when using larger GW detector networks that include more than the three detectors considered here. 
 
\section{Acknowledgements}\label{sec:acknowledgements}
The work of EF and VM was in part supported by NSF grants PHY-2110238 and PHY-1806630.  The authors are grateful for computational resources provided by the LIGO Laboratory and supported by NSF Grants PHY-0757058 and PHY-0823459. This material is based upon work supported by NSF's LIGO Laboratory which is a major facility fully funded by the National Science Foundation.

\bibliography{refs}
\bibliographystyle{apsrev4-1}
%\bibliographystyle{plain}

%20-500, 2/3 fkeep, HL or HLV, show how departures from process don't help 
%departure 1: use HLV, then can you use different fkeep?
%departure 2: use different frequency ranges
%departure 3: use different alpha (backburner)
%try weaker signals for two point sources, investigate lmax=5, smaller frequency bands
%combine HL and HLV spot size graphs
% 20-200Hz
%4 SNR maps of 2points
%intro pages: GWs and LIGO 2-3
    % SHD search 2-3
    % O3 SHD 3
    % angular resolution 5
%O3 results in oral exam paper
\end{document}